\documentclass[final,12pt]{elsarticle}

\usepackage{graphicx}

\usepackage{amssymb}
\usepackage{amsmath}
\usepackage{color}
\usepackage{comment}

\definecolor{darkblue}{rgb}{0,0,.6}
\definecolor{darkred}{rgb}{.6,0,0}
\definecolor{darkgreen}{rgb}{0,.6,0}
\definecolor{red}{rgb}{.98,0,0}

\def\ssmall{\fontsize{8pt}{2pt}\selectfont}

\usepackage{fancybox}

\PassOptionsToPackage{normalem}{ulem}
\usepackage{ulem}
\usepackage{subfig}
\usepackage{multirow}
\usepackage{multicol}

\usepackage{colortbl}
\definecolor{lightgray}{gray}{0.8}

\newcounter{bla}

 \usepackage{hyperref}
\usepackage{listings}
\lstnewenvironment{cpplisting}[1][]
  {\lstset{language=C++,numbers=right,#1}}{}

\lstnewenvironment{bashlisting}[1][]
  {\lstset{language=bash,numbers=right,#1}}{}

\lstnewenvironment{outputlisting}[1][]
  {\lstset{language=html,numbers=right,#1}}{}

\journal{Computer Physics Communications}

\begin{document}

\newcommand{\ME}{\begin{tt}Maxent\end{tt}}

\begin{frontmatter}

\title{LIBAMI: Implementation of Algorithmic Matsubara Integration}

\author[a]{Hossam Elazab}
\author[a]{B. D. E. McNiven}
\author[a]{J. P. F. LeBlanc\corref{author}}

\cortext[author] {Corresponding author.\\\textit{E-mail address:} jleblanc@mun.ca}
\address[a]{Department of Physics and Physical Oceanography, Memorial University of Newfoundland, St. John's, Newfoundland \& Labrador, A1B 3X7, Canada}

\begin{abstract}
We present libami, a lightweight implementation of algorithmic Matsubara integration (AMI) written in C\texttt{++}.  AMI is a tool for analytically resolving the sequence of nested Matsubara integrals that arise in virtually all Feynman perturbative expansions.  
\end{abstract}

\begin{keyword}
Algorithmic Matsubara Integration \sep Feynman Diagrams \sep Diagrammatic Monte Carlo

\end{keyword}

\end{frontmatter}



{\bf{PROGRAM SUMMARY}}

\begin{small}
\noindent
{\em Program Title:}  libami                                        \\
{\em CPC Library link to program files:} \\
{\em Developer's repository link:} \url{https://github.com/jpfleblanc/libami} \\
{\em Code Ocean capsule:} (to be added by Technical Editor)\\
{\em Licensing provisions:} GPLv3\\
{\em Programming language:}  \texttt{C++}                                 \\
{\em Nature of problem:}\\
Perturbative expansions in condensed matter systems are formulated on the imaginary frequency/time axis and are often represented as a series of Feynman diagrams, which involve a sequence of nested integrals/summations over internal Matsubara indices as well as other internal variables. \\
{\em Solution method:}\\
 \texttt{libami} provides a minimal framework to symbolically generate and store the analytic solution to the temporal Matsubara sums through repeated application of multipole residue theorems.  The solution can be applied to any  frequency-independent interaction expansion. The analytic solution once generated is valid in any dimensionality, with any dispersion at arbitrary temperature. \\
{\em Additional comments including restrictions and unusual features:}\\
Requires \texttt{C++11} standard.
   \\

\end{small}

\section{Introduction}
The Feynman diagram approach to many-body perturbation theory is widely used within the fields of high-energy and condensed matter physics.\cite{mahan}  Within condensed matter physics, many-body perturbative expansions form the foundation for understanding model Hamiltonians and experimentally realizable observables.  Beyond model systems, Feynman diagrams are useful in extensions of ab-initio methods based on various $GW$ approximation schemes that find wide applicability within the fields of both physics and quantum chemistry.\cite{gunnarsson:gwreview} While chemistry methods tend to focus on ground state properties of molecules and materials, the central interest for physicists are the details of finite-temperature phase transitions and dynamical properties of correlated electron systems.\cite{Schaefer:2020} 

With these targets in mind, most numerical approaches to finite temperature diagrammatics employ the so-called imaginary-time or Matsubara formalism.\cite{mahan} The results of such calculations (Green's functions, self energies, and two-particle correlation functions) are obtained only in imaginary time or frequency and the data must then be numerically inverted to obtain real frequency results, a process that is fundamentally ill-posed.\cite{jarrell:maxent,Levy2016} 
As a result, computational methods like $GW$ truncate the expansion at the lowest order (zeroth order) where analytic results are easily available, allowing one access to quantities on the real-frequency axis.\cite{kresse:2007,kresse:2018}  
Accessing the perturbative expansion to higher order for the real-frequency axis is central to improving upon those methods. 
Methods do exist that can sum the perturbation expansion such as 
Diagrammatic Monte Carlo \cite{vanhoucke} or its extensions such as connected-determinant Monte Carlo,\cite{rossi2017determinant,fedor:2020} that provide access to higher expansion order.  However, these methods again operate on the Matsubara axis and cannot resolve the analytic continuation issue. 

Recent work\cite{AMI} provides a way around analytic continuation known as algorithmic Matsubara integration (AMI). The key realization leading to the advent of AMI is that in principle, the procedure for evaluating the internal Matsubara sums is known, but doing so for all but the simplest diagrams produces analytic expressions with numbers of terms growing sub-factorially with expansion order. 
One can in principle obtain analytic representations for certain expansions but the resulting expressions are so complicated that the utility of such results is compromised since there is a larger barrier for a user to apply the result.\cite{jaksa:analytic} 

AMI follows the mindset of tools such as automatic differentiation\cite{auto_diff} where the user is intentionally never exposed to the resulting mathematical expressions, and where those expressions are never themselves hardcoded.  Instead, the user need only provide the structure of the integrand of interest, and AMI provides the tools to first construct the symbolic analytic representation of the temporal Matsubara integrals, and then provide the tools to evaluate the resulting expressions for choices of internal/external variables. 
In this work, we describe the fundamentals of AMI and present the compact, virtually dependence-free library \emph{libami}. 

\section{Algorithm and Implementation}
In this section, we paraphrase the AMI methodology first presented in Ref.\cite{AMI} and its supplementary materials as well as its extensions presented in  Ref.\cite{GIT}. 

\subsection{Explanation of the problem}
The problem we wish to solve is the integration over internal variables of a Feynman diagram.  Making no assumptions about the topology of the diagram,
the general form of a Feynman integrand can be written as
\begin{align}
&\frac{1}{\beta^n}\prod\limits_{i}^{n_v}V(q_i)\sum\limits_{\{k_n\}} \sum \limits _{\{\nu_n\}}\prod
\limits_{j=1}^N G^j(\epsilon ^j, X^j ) = \prod\limits_{i}^{n_v}V(q_i)\sum\limits_{\{k_n\}} I^{(n)}, \\
& I^{(n)}=\frac{1}{\beta^n}\sum \limits _{\{\nu_n\}}\prod
\limits_{j=1}^N G^j(\epsilon ^j, X^j ), \label{eqn:goal}
\end{align}
where $n_v$ is the order or the number of interaction lines with amplitude $V(q_i)$ of the diagram. $n$ 
is the number of summations over Matsubara frequencies $\{\nu_n\}$  and 
internal momenta $\{k_n\}$, and $N$
is the number of internal lines representing bare Green's functions $G(\epsilon,X)$.
We choose a sign convention such that the bare Green's function of the $j$th internal line is
\begin{eqnarray}
G^j(\epsilon^j, X^j) =  \frac{1}{ X^j + \epsilon^j},\label{eqn:Green_Function}
\end{eqnarray}  
where $X^j$ is the frequency and $\epsilon^j = -\epsilon(k_j)$ is the negative of the free particle dispersion where $k_j$ is the momentum of the $j$th Green's function. 
Constraints derived from energy and momentum conservation at each vertex 
allow us to express the parameters of each free-propagator line as linear combinations of internal $\{\nu_n, {k}_n\}$ and external $\{ \nu_\gamma, k_\gamma\}$ frequencies and momenta, where $k_j  =  \sum_{\ell=1}^m \alpha_{\ell}^j k_{\ell} $, $X^j   =   \sum_{\ell=1}^{m}i\alpha_\ell^j \nu_{\ell}$.

While it is not obvious, the coefficients $\alpha_\ell^j$ are integers which for most problems need only have three possible values:  zero, plus one, or 
minus one. 
This allows us to represent $G^j$ symbolically in a combined structure
\begin{eqnarray}\label{E: Green_Function_Array}
G^j(X^j) \to [\epsilon^j, \vec \alpha^j],
\end{eqnarray} 
where $\vec \alpha^j = (\alpha_1^{j}, ..., \alpha_{m}^{j})$. We emphasize that $\epsilon^j$ has absorbed a minus sign that normally appears in the denominator for Fermionic Green's functions. In the simplest incarnation of AMI, each $\epsilon^j$ energy is a single complex number.    In this implementation however, we redefine each $\epsilon^j$ to be a linear combination of all \emph{initial} energies, $\epsilon_0$, that appear in the starting integrand as $\epsilon^j=\sum\limits_{j^\prime} a_{j^\prime}\epsilon_0^{j^\prime}$, with $a_{j^\prime}$ again typically adopting values of $\pm1$ or 0. This will be discussed in detail in the implementation section and examples will be provided. 

Eqn.~(\ref{E: Green_Function_Array}) is the key symbolic object we will need to manipulate.   Given the array representation of each individual $G^j$, we can construct a nested array of such objects to represent the product of $G^j$ which appears in Eq.~(\ref{eqn:goal}),
\begin{eqnarray}\label{E: Only_One_New_General_Sum_3}
\prod_{j=1} ^ N G^j(\epsilon^j, X^j) \to \bigg [[\epsilon^1, \vec \alpha^1]; [\epsilon^2, \vec \alpha^2]; ...; [\epsilon^N, \vec \alpha^N] \bigg]. \label{eqn:gprod}
\end{eqnarray}
This representation 
carries all the information we need to compute
the summations in Eq.~(\ref{eqn:goal}).

To begin the algorithm, we subdivide the original problem to the summation over a single frequency $\nu_p$, and the remaining frequencies $\nu_n\neq \nu_p$,
\begin{align}
&I^{(n)}=\sum_{ \{\nu_n\}, \nu_n\neq \nu_p } I_p, \label{eqn:onesum} \\
&I_p = \sum_{\nu_p} \prod_{j=1} ^ N G^j(\epsilon^j, X^j_m).\label{E:Only_One_New_General_Sum}
\end{align}  
The result of each Matsubara summation  over the fermionic frequency $\nu_p$ is given by the residue theorem,
\begin{eqnarray}\label{E: Res_Th}
\sum_{\nu_p} h(i\nu_p) = \beta \sum_{z_p} f(z_p){\rm Res}[h(z)]_{z_p},
\end{eqnarray}
where $f(z)$ is the Fermi-distribution function
and $z_p$ are the poles of $h(z)$. 
Central to computing Eq.~(\ref{E:Only_One_New_General_Sum}) is the identification of the set of poles of the Green's functions.  The pole itself therefore contains the same information as the initial Green's function, but with respect to a specific internal frequency index.  It can therefore be stored in a structure that is nearly identical to the Green's functions themselves. 

The pole of the  $j$th Green's function with respect to the frequency $\nu_p$ 
exists so long as the coefficient $\alpha_p^j$  is non-zero, and is given by 
\begin{eqnarray}\label{E: Poles_one_h(z)}
z_p ^{(j)} = -\alpha_{p} ^{j} ( -\epsilon^j + \sum_{\ell=1, \ell\neq p} ^m i\alpha_\ell^{j}\nu_\ell)  \ \ \ \ \text{for} \ \ \ \alpha_{p}^{j} \neq 0 .
\end{eqnarray}
The number of simple poles for $\nu_p$ is
$r_p = \sum _{j=i}^{N} |\alpha_p^i|$, which appear in a number, $r_p$, of $N$ total 
Green's functions in the product of Eq.~(\ref{E:Only_One_New_General_Sum}). 
If the exact pole appears more than once, the pole has a multiplicity, $M>1$, which gives rise to additional complications and therefore the general case for poles with multiplicity $M$ must be used. If $h(z)$ has a pole of order $M$ at $z=z_0$, then the residue is given by 
\begin{eqnarray}\label{E: Res_mul}
Res[h(z_0)] = \frac {1} {(M-1)!} \lim_{z\to z_0} \frac {d^{M-1}}{dz^{M-1}} \bigg \{(z-z_0)^Mh(z)\bigg \}. \nonumber \\
\end{eqnarray}

%
In the case of simple poles, the Fermi function is evaluated as
\begin{eqnarray}\label{E: Fermi_Simpl_diff}
f(z_p^{(i_\ell)}) =  \frac{1}{\sigma \exp(-\beta \alpha_p^{i_\ell}\epsilon^{i_\ell}) + 1},
\end{eqnarray}  
where  the additional negative in the exponential argument is due to the sign convention of the dispersion appearing in the Green's function and $\sigma$ is a sign given by
\begin{align}\label{E: Fermi_sign}
&\sigma(z_p^{i_\ell}) = \exp (i\beta\sum_{\ell\neq p} \alpha_\ell^{i_\ell} \nu_\ell),
\end{align}  
that is, $\sigma = -1$ if there are an odd number of fermionic frequencies in the sum over
$\ell$, otherwise $\sigma = 1$.
Therefore, $f(z_p^{(i_\ell)})$ is independent of Matsubara frequencies and only depends on the real energy dispersion, though its character might switch from fermionic to bosonic. 
In the case of multipoles, derivatives of Fermi/Bose functions will arise (see \ref{app:der} ).

We make use of this result to 
calculate {\em all} of the summations in  Eq.~(\ref{eqn:goal}) 
using a recursive procedure, since the problem after each integration step has the same analytic form in our array notation.

Since multipole cases incur derivatives, we employ the method of automatic (or algorithmic) differentiation to analytically evaluate arbitrary order derivatives, as this procedure requires only knowledge of the first derivative and repeated application of chain rules.\cite{auto_diff} The first derivative with respect to $i\nu_p$ of the multiplication of $N$ Green's function is given via the chain rule as
\begin{eqnarray}\label{E: Derivation_1_func}
\frac{d}{d(i\nu_p)}(\prod_{j=1} ^ N G^j(\epsilon^j, X^j_m)) = \sum_{i=1}^{N} \frac{dG^i}{d(i\nu_p)} \prod_{j\neq i} G^j(\epsilon^j, X^j_m). \nonumber \\
\end{eqnarray}  
The first derivative of one of the Green's functions with respect to $i\nu_p$ in the array representation can then be performed by returning two Green's functions,
\begin{eqnarray}\label{E: Derivation_G_array}
\frac{dG^i(\epsilon^i, X_m^i)}{d(i\nu_p)} \to \bigg [ [\epsilon^i,X_m^i]; [-\alpha_p^i \epsilon^i, \alpha_p^i X_m^i] \bigg].
\end{eqnarray}  
The $(M-1)$th order derivative can be computed by iterating Eq. (\ref {E: Derivation_1_func}).  We therefore are able to express the residue for poles of $i\nu_p$ with any multiplicity using our symbolic representation.   There are two significant differences from the simple-pole case.  Firstly, the presence of multiple poles results in factors of $\pm \frac{1}{(M-1)!}$ that must be stored. Secondly, the chain rule results in additional terms that themselves result in higher order poles leading to factorial growth in number of analytic terms. The derivative also contains Fermi functions that add an additional step to the chain rules via
\begin{eqnarray}\label{E: residue_derivative}
\frac{d}{dz}\left(f(z)\prod_{j=1} ^ N G^j \right) = \frac{d f(z)}{dz} \prod_{j=1} ^ N G^j + f(z)\frac{d}{dz}\left(\prod_{j=1} ^ N G^j\right) .\nonumber \\
\end{eqnarray}  

We see that one must therefore track the Fermi derivative order.  This can be stored with each individual pole and applied upon evaluation of the result.

\subsection{Implementation}

  \begin{figure*}[t]
\centering
  \includegraphics[width=\linewidth]{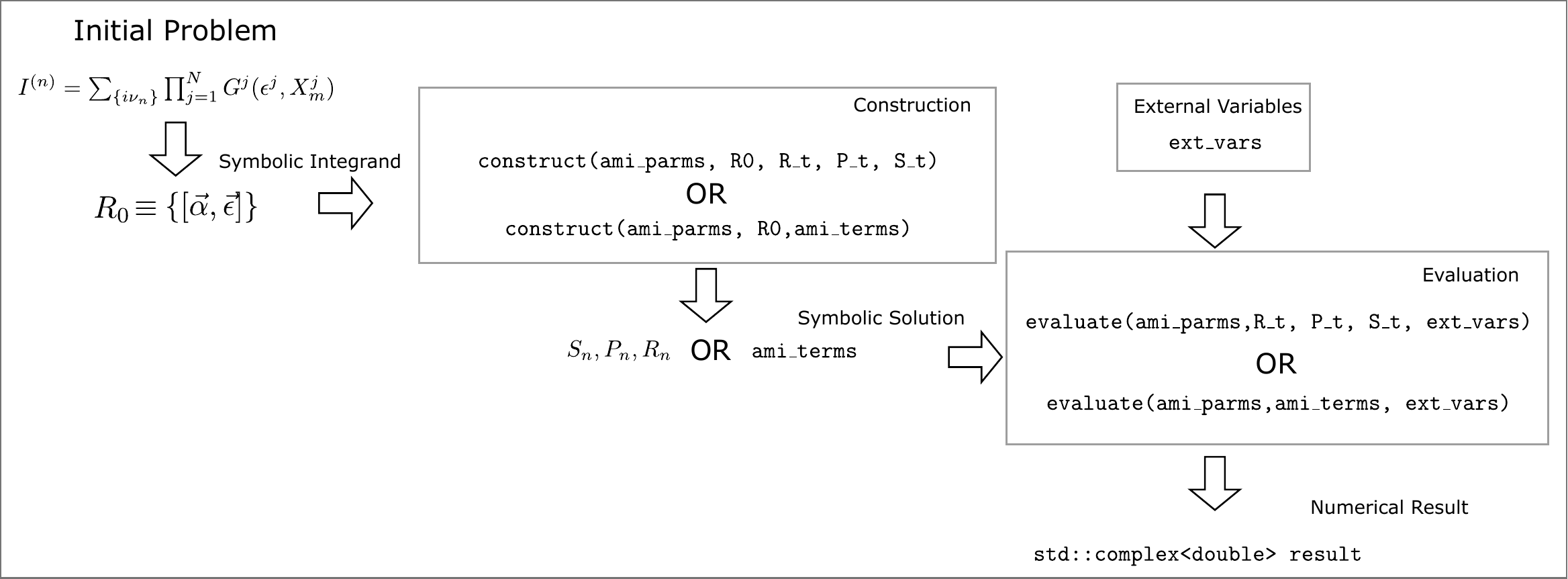} 
 \caption{Schematic describing the workflow of the AMI routine.}
 \label{fig:workflow}
 \end{figure*}

To solve the integrals in Eq.~(\ref{eqn:goal}), the AMI routine is broken into two main components depicted in Fig \ref{fig:workflow}. 
These are: 1) the construction of a symbolic integrand 2) the evaluation of the integrand for a set of external/internal parameters.

\subsubsection{Construction}

The user must first generate a representation of the integrand of interest via Eq.~(\ref{eqn:gprod}).  
Each Green's function has a structure of type \texttt{g\_struct} and these are then stored in the object \emph{R0} that is defined as a 
\texttt{g\_prod\_t}.  The  \texttt{g\_prod\_t} is
simply a vector of \texttt{g\_struct} objects that is interpreted as a product upon evaluation. 
The analytic solution to the Matsubara sums can then be determined by two independent methods. We note that both methods of handling \emph{R0} produce the same analytic solution, but differ in overall complexity of evaluation both in computational and memory resources. We first introduce the fastest method of producing the analytic solution using what we call the $S$, $P$, and $R$ arrays. This is the construction defined in the initial AMI publication\cite{AMI} and we elaborate in \ref{app:ami}. These arrays, stored as \texttt{std::vector} containers, track the sign $(S)$ of each residue at the $p^{th}$ pole, the $p^{th}$ pole itself $(P)$, and the $p^{th}$ Green's function configuration $(R)$.  Subsequently, at evaluation time the S, P and R are manipulated to produce the final integrand. The downside of the $S$,$P$,$R$ approach is that one cannot look at the numerator of the analytic solution on a term-by-term basis.  Rather, one obtains all the term numerators in a highly factorized manner.  In order to extract specific analytic terms, one must use the \texttt{terms} object to do so. While it is slower in comparison, it also allows for much more intuitive manipulation, making it more user friendly for one who is not familiar with the formal AMI procedure of Ref.~\cite{AMI}. The latter approach is essential if the numerator of each individual term is required as is often the case for variance reduction integration techniques.  

In the term-by-term construction approach, each term of the resulting expression is defined by three pieces: 1) an overall prefactor, 2) a set of poles (\texttt{std::vector<pole\_struct>}) representing a product of Fermi/Bose functions in the numerator, and 3) a product of Green's functions (\texttt{g\_prod\_t}) 
representing the denominator.  Each term is defined as a structure containing these objects and an entire AMI integrand is then just a vector of such terms.  In comparison to the SPR construction/evaluation, the \texttt{terms} representation is always slower since at evaluation time one would see many duplicate poles that are then repeatedly evaluated.  
This is avoided in the SPR construction since each pole and prefactor will only appear once and be appropriately distributed to all terms. 

As shown in Fig.~\ref{fig:workflow}, the two construction methods share the same construct function name and differ in the number and types of arguments.  There is currently no mechanism to convert from $S,P,R$ arrays to an \texttt{terms} object.  Doing so is not only complicated but also nearly identical to simply reconstructing the term-by-term result.  Since the computational expense of construction is low, this is not problematic.

\subsubsection{Evaluation}
With the analytic solution constructed, evaluating the expressions now requires the external variables specified by the user to produce the numeric solution. For most problems, the required external variables include the inverse temperature ($\beta$), the complex chemical potential and external frequency,  as well as the specific k-space  coordinates ($k_i$) and the integrand's dimensionality.
However, from the perspective of AMI most of these external parameters are problem specific.  The AMI routine only needs a set of energies appearing in $R0$ as well as in the external frequencies.  It is therefore left to the user to define their dispersion and populate the energies accordingly.  Thus, the only inputs to the AMI evaluation are:
\begin{itemize}
    \item \texttt{frequency\_t} - \texttt{std::vector< std::complex<double>>} that represents the vector $(\nu_1,\nu_2,...,\nu_{n+m})$, the list of $n$ internal frequencies and $m$ external frequencies. 
    \item \texttt{energy\_t} - \texttt{std::vector< std::complex<double>>} that represents the vector containing the negatives of energies appearing in Eq. \ref{eqn:goal} and \ref{eqn:Green_Function}, $(-\epsilon_1,-\epsilon_2,...,-\epsilon_N)$.  Note the storage of the negative of the energy is a convention. 
\end{itemize}

\subsection{Current Limitations}
We mention here some limitations that will be addressed in a future release with priority dependent on user feedback. 

\subsubsection{Fundamental}
\begin{enumerate}
    \item All independent internal frequencies for summation are Fermionic.
    \item Only providing control of single external frequency statistics (Fermi/Bose).
    \item Multi-leg diagrams are largely untested.
    \item No catch for spurious multi-poles.
\end{enumerate}

Items (1) and (2) above limit functionality to diagrammatic expansions of Fermionic operators. With this restriction, \texttt{libami} can be applied to any correlated electron system.  It cannot be applied, in its current form, to mixtures of Fermionic and Bosonic particles (electron-phonon interactions) nor purely Bosonic systems.  Resolving this limitation results in somewhat less performant evaluation and we have therefore removed it from this release. 

The exception to the above is the case where the external lines depend on a single Bosonic Matsubara frequency, as is the case for two-particle susceptibilities that are commonly of interest, see the Examples section below.

Spurious poles, mentioned in item (4) above, are poles that occur when denominators of multiple Green's functions with distinct arguments to their energies (momenta) happen to return the same value.  This is a very common occurrence and is exacerbated when performing the internal momentum summations if the momenta are evaluated on a fixed grid.  It is recommended that the momenta be evaluated at random to avoid the coincidental overlap of two symbolically distinct poles.  Alternatively, if one needs to evaluate these spurious points, it can be accomplished by reconstructing the integrand to symbolically reflect the equivalence of those energies (see the fourth-order multi-pole example).

\subsubsection{Extensions}
\begin{enumerate}
    \item No graph framework to generate diagrams, labels and \texttt{libami} input.
\end{enumerate}
Automating the generation of the AMI input starting from operators or from a particular diagrammatic expansion can be accomplished,\cite{GIT,AMI:spin,mcniven:2021} but is not intended to become a part of \texttt{libami}.  Instead, the necessary components for full automation will be developed using libami as the backend.

\subsection{Optional Optimizations}
Once constructed, either in $S,P,R$ or \texttt{terms} form, the integrand can be preprocessed to reduce computational expense of the evaluation stage.  Even for very complicated diagrams with hundreds or thousands of terms, one will find that after evaluating the internal Matsubara sums, there are only a handful of unique Green's functions. It is therefore preferable to identify the unique components and evaluate each only once.  

The factorized form can be found via 
\begin{cpplisting}
  void AmiBase::factorize_terms(terms &terms, g_prod_t &unique_g, R_ref_t &Rref,ref_eval_t &Eval_list)
  \end{cpplisting}
  in the case of the \texttt{terms} form, and via
\begin{cpplisting}
  void AmiBase::factorize_Rn(Ri_t &Rn, g_prod_t &unique_g, R_ref_t &Rref,ref_eval_t &Eval_list)
\end{cpplisting}
for the $S,P,R$ form. Regardless of which is used, the final three arguments can then be passed to the evaluate functions by simply appending them to the argument list of \texttt{evaluate}.  Explicit examples are provided. 

\section{Installation, Documentation and Tests}

\subsection{Required Libraries}
\texttt{libami} uses only standard libraries within the \texttt{c++11} standard.  This requires \texttt{GCC>=4.8.1} or \texttt{INTELC++ >= 15.0} or \texttt{clang >= 3.3}.  It is recommended however to use more up-to-date compilers due to small but noticeable performance improvements. 

\subsection{Compilation, Documentation and Tests}
Once obtained, the library can be compiled via \texttt{cmake} from the \texttt{libami} directory via
\begin{bashlisting}
$ mkdir build
$ cd build
$ cmake -DCMAKE_INSTALL_PREFIX=/where/to/install/libami ..
$ make install
\end{bashlisting}

Tests are disabled by default but are enabled when compiled in release mode and can be enabled manually via 
\begin{bashlisting}
$ cmake -DCMAKE_INSTALL_PREFIX=/where/to/install/libami -DTEST=ON ..
$ make test
\end{bashlisting} 

The testing framework utilizes built in \texttt{cmake} testing functionality that requires \texttt{cmake} version $>=$3.18.  The code can be compiled with older versions of \texttt{cmake} but the test compilation will fail. 

The code also has some additional documentation that can be compiled.  Doing so requires \texttt{doxygen} as well as the \texttt{Sphinx} package in \texttt{python3}.  With these dependencies met the command:
\begin{bashlisting}
$ cmake -DCMAKE_INSTALL_PREFIX=/where/to/install/libami -DBUILD_DOC=ON ..
\end{bashlisting}
should compile and build documentation.  Typically this is not necessary as HTML versions of the documentation will be made available through the \texttt{git} repository.

\section{Examples}
Often it is the case that explicit examples have more utility for users than documentation.  This is the case for \texttt{libami} since it is intended that users should not have to delve into the library, but instead primarily interact with the code through just a few functions.  We provide a handful of specific examples, where each case the procedure is the same:
\begin{enumerate}
    \item Define integrand $R0$.
    \item Construct AMI solution.
    \item Define internal parameters and evaluate.
\end{enumerate}

For each example, we sequentially evaluate with the $S,P,R$ formulation, the \texttt{terms} formulation, as well as the factorized forms. 

\subsection{Example 1: Second Order Self-energy}
This example is defined by the \texttt{example2()} function in the examples directory \texttt{src} files.
\subsubsection{Define Integrand}
The starting integrand for the second order self energy is given by
\begin{equation}\label{eqn:r0}
    \Sigma(\nu_3,k_{ext})=\frac{1}{\nu_1-\epsilon_{k_1}}\frac{1}{\nu_2-\epsilon_{k_2}}\frac{1}{-\nu_1+\nu_2+\nu_3-\epsilon_{k_3}},
\end{equation}
where for momentum conservation $k_3=k_{ext}-k_1+k_2$. The conversion to a symbolic code requires of course, assigning labels $i$ to the various frequencies $\nu_i$.  It does not matter how the internal labels are assigned, but it is essential that the final index ($\nu_3$) be the external line.  The AMI construct function performs \texttt{NINT} nested integrals from $i=1\to$\texttt{NINT}.  

To begin, we store the frequencies in each denominator in a set of \texttt{alpha\_t} vectors,
\begin{cpplisting}
AmiBase::alpha_t alpha_1={1,0,0};
AmiBase::alpha_t alpha_2={0,1,0};
AmiBase::alpha_t alpha_3={-1,1,1};
\end{cpplisting}
and the energy denominators as,
\begin{cpplisting}
AmiBase::epsilon_t epsilon_1={1,0,0};
AmiBase::epsilon_t epsilon_2={0,1,0};
AmiBase::epsilon_t epsilon_3={0,0,1};
\end{cpplisting}
Here, each \texttt{epsilon\_i} is a placeholder for the negative of $\epsilon_i$.  While in this case the geometry of the \texttt{alpha\_t} and \texttt{epsilon\_t} are the same, this typically will not be true for arbitrary diagram choices.  Specifically, the length of each \texttt{alpha\_t} is related to the order of the diagram, which in this case has length $\ell=M+1$ for an $M$th order diagram.  The length of each \texttt{epsilon\_t} is the number of Green's functions given by $2M-1$.  One may note immediately the \texttt{epsilon\_t} when listed form an $N\times N$ identity matrix, which is always the starting point for virtually any problem.  If it is known in advance that two of the energies are equivalent, then they can subsequently be set equal. 
We then define the starting integrand via:
\begin{cpplisting}
AmiBase::g_struct g1(epsilon_1,alpha_1);
AmiBase::g_struct g2(epsilon_2,alpha_2);
AmiBase::g_struct g3(epsilon_3,alpha_3);

AmiBase::g_prod_t R0={g1,g2,g3};
\end{cpplisting}
Each \texttt{g\_struct} has an \texttt{alpha\_t} and \texttt{epsilon\_t} that can be assigned/accessed via \texttt{g\_struct.alpha\_} and \texttt{g\_struct.eps\_}, or, alternatively assigned via the constructor as in lines $1\to3$ above. The resulting integrand is stored as a \texttt{g\_prod\_t} which is a vector of Green's functions.
The ordering of the three Green's functions in \texttt{R0} will not impact the result.

\subsubsection{Construct Solution}

Once the integrand is defined, the construction requires three steps. i) Instantiate the class and define containers to store the solutions. ii) Define the integration parameters \texttt{AmiBase::ami\_parms}. iii) construct the solution. 

These can be accomplished with the following lines
\begin{cpplisting}
AmiBase ami;
AmiBase::S_t S_array;
AmiBase::P_t P_array;
AmiBase::R_t R_array;

double E_REG=0; 
int N_INT=2;  
AmiBase::ami_parms test_amiparms(N_INT, E_REG);

ami.construct(test_amiparms, R0, R_array, P_array, S_array); 
\end{cpplisting}

Part (i) is accomplished in lines $1\to4$, (ii) in lines $6\to8$ and (iii) in line $10$.  The \texttt{ami\_parms} structure requires an energy regulator \texttt{E\_REG} to be used at the evaluation stage (typically not needed and set to zero in this case), and the number of Matsubara sums to perform, \texttt{N\_INT}. 
Line $10$ above contains the call to the \texttt{construct} function which combines these elements and populates the S,P,R objects, \texttt{S\_t}, \texttt{P\_t}, and \texttt{R\_t}, respectively.

The alternate term-by-term construction can be obtained by replacing lines $2\to4$ with:
\begin{cpplisting}
AmiBase::terms amiterms;
\end{cpplisting}
and line $10$ with:
\begin{cpplisting}
  ami.construct(test_amiparms, R0, amiterms);
\end{cpplisting}

\subsubsection{Evaluate}

While the construction phase works with integer arrays that represent symbols in the starting integrand, in order to evaluate the result, one would need to specify values for each $\epsilon_i$ and $\nu_i$ symbol in Eq.~\ref{eqn:r0} and these are stored in types \texttt{energy\_t} and \texttt{frequency\_t} respectively.  While the \texttt{energy\_t} and \texttt{frequency\_t} have the same geometry as \texttt{epsilon\_t} and \texttt{alpha\_t}, they contain the actual values that are of type \texttt{std::complex<double>}.  

For this example, we choose a particular set of values:
\begin{cpplisting}
AmiBase::energy_t energy={-4,0.1,-1};

AmiBase::frequency_t frequency;
for(int i=0;i<2;i++){ frequency.push_back(std::complex<double>(0,0));}
frequency.push_back(std::complex<double>(0,M_PI));
\end{cpplisting}
Being careful of the energy sign convention, the \texttt{energy\_t} has absorbed the negative signs in the denominators of Eq.~\ref{eqn:r0}.  This is therefore equivalent to the integrand
\begin{equation}\label{eqn:exeval}
    \Sigma(\nu_3,k_{ext})=\frac{1}{\nu_1-4}\frac{1}{\nu_2+0.1}\frac{1}{-\nu_1+\nu_2+\nu_3-1}.
\end{equation}

The values in the \texttt{energy\_t} and \texttt{frequency\_t} are considered `external' to AMI in the sense that they are constant with respect to the Matsubara summation.  The exception is of course the first entries of \texttt{frequency\_t}.  We see that there are empty placeholders of zero for $\nu_1$ and $\nu_2$.  After integration $\nu_1$ and $\nu_2$ will not appear in the AMI result, and so these values do not matter.  However, the evaluate function allows that one might perform only \emph{some} (or even none) of the Matsubara sums, in which case specific values can be applied.  Here the only frequency that is relevant is the final one, $\nu_3=i\pi$.  This is the evaluation of the self energy for a Matsubara frequency $i\pi/\beta(2n+1)$.  We will set $\beta=1$ so that this represents the zeroth Matsubara frequency. Alternatively, we are free to assign $\nu_3$ any value, real or imaginary, such as the usual analytic continuation $\nu_3 \to \omega+i\Gamma$ where $\omega$ is a real frequency and $\Gamma \to 0^+$ is a small positive scattering rate. 

These external objects are combined into a single structure 
of evaluation variables \texttt{AmiBase::ami\_vars} via:
\begin{cpplisting}
AmiBase::ami_vars external(energy, frequency,BETA);
\end{cpplisting}
Finally the actual numerical evaluation is done via:
\begin{cpplisting}
  std::complex<double> calc_result=ami.evaluate(test_amiparms,R_array, P_array, S_array,  external);	
\end{cpplisting}
which returns an \texttt{std::complex<double>} result. 

Similar to the construction phase, one can evaluate the term-by-term solution via a similar exchange of the \{\texttt{R\_array, P\_array, S\_array}\} arguments with the AMI \texttt{terms} struct:
\begin{cpplisting}
  std::complex<double> term_val=ami.evaluate(test_amiparms, amiterms, external);
\end{cpplisting}

\subsection{Example 2: Bosonic External Frequencies}
As a second example, defined in the \texttt{example1\_bose()} function, we outline the procedure for generating the solution to the standard particle-hole bubble.  This problem involves only a single Matsubara sum, but requires that the external frequency be Bosonic.
The setup of this example is identical to the setup to the previous example.  The key change being the constructor call for \texttt{AmiBase::ami\_parms}.  Specifically, an additional parameter should be given:
\begin{cpplisting}
AmiBase::ami_parms test_amiparms(N_INT, E_REG,1);  
\end{cpplisting}
which is equivalent to setting \texttt{test\_amiparms.graph\_type=1}.  This flag tells the \texttt{evaluate} function to handle the external frequency as Bosonic while the default value is Fermionic (\texttt{test\_amiparms.graph\_type=0}). 

Furthermore, a noteworthy point is our current AMI implementation \emph{requires} that all internal $\nu_i$ are Fermionic Matsubara frequencies.  Therefore, the diagram must be labelled such that independent internal labels are assigned to Green's functions and \emph{not} to interaction lines.  Although this is a limitation, it is a massive simplification at run-time that avoids repeated checks of statistics flavor. 

\subsection{Further Examples}
We include two additional examples, \texttt{example4()} and \texttt{example9()}, for a fourth order and ninth order problem, respectively.  The fourth order problem illustrates a multipole problem, while the ninth order case demonstrates an extreme example that pushes the limitations of the current implementation.  By ninth order, the result of the AMI procedure is on the order of $1\times10^5$ terms, and a single evaluation is on the order of 1~second.  While this is a perfectly reasonable timescale for a single evaluation, the integration over the remaining internal variables typically would require millions of such samples.

\section{Benchmarks}\label{sec:benchmarks}
It is straightforward to use \texttt{libami} to evaluate quantities for both real and Matsubara frequencies.  In the latter case, the problem is then identical to canonical Diagrammatic Monte Carlo methods\cite{vanhoucke,signblessing}.  However, in those methods, the integration space is substantially larger, whether evaluated in imaginary time, $\tau$, or in frequency space. While in principle Monte Carlo methods converge regardless of the problems dimensionality, it is extremely costly to sample large integration spaces if the integrand itself is sharply peaked, as is especially the case in the real frequency domain.  

\begin{figure}
    \centering
    \includegraphics[width=0.8\linewidth]{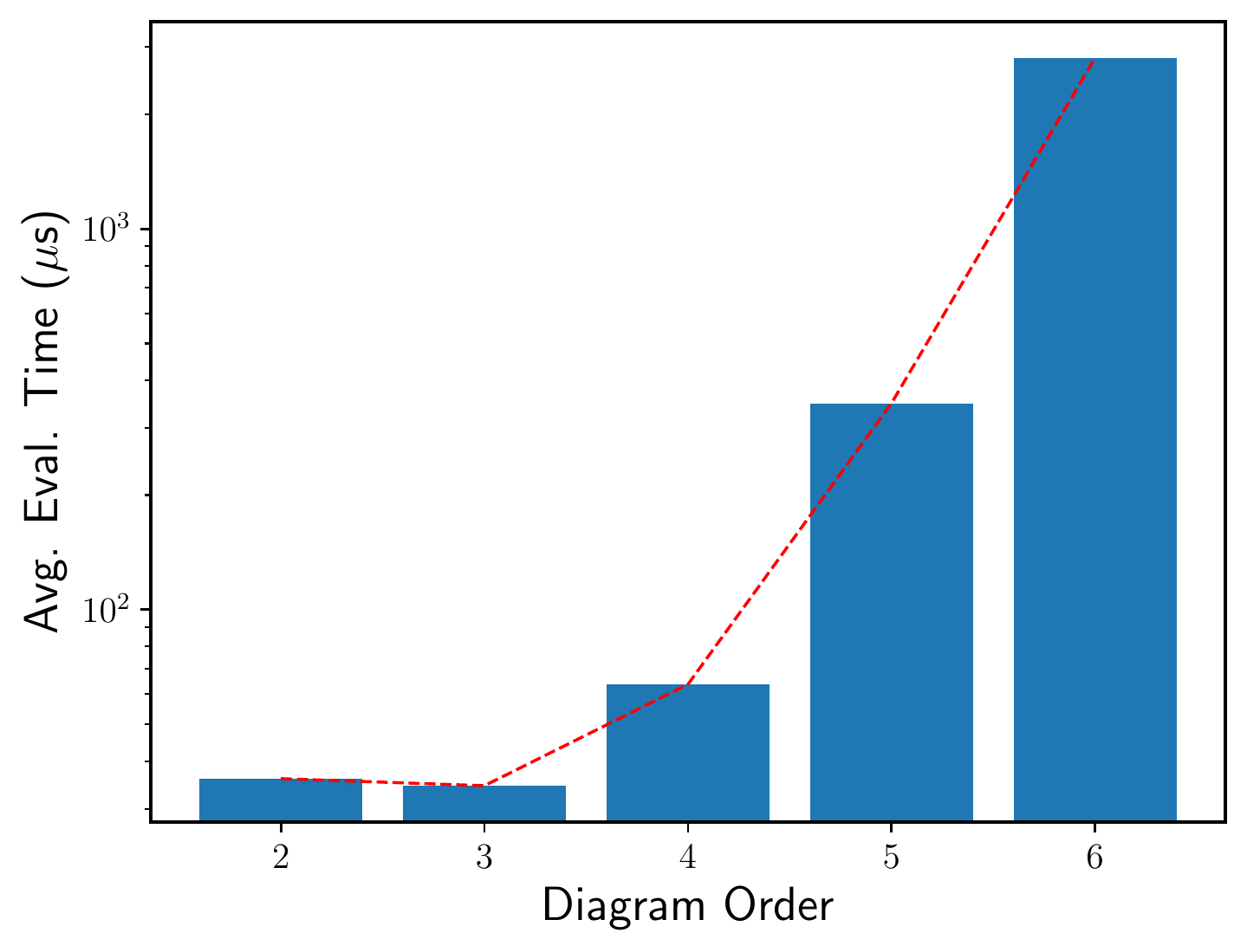}
    \caption{
    \label{fig:hist} Average time for evaluation (in $\mu$s) per diagram for evaluation in S,P,R format for self-energy diagrams for a random choice of internal parameters on a single modern 3.5~GHz CPU.}
\end{figure}

In order to provide a scale associated with the application and evaluation of \texttt{libami}, the examples contain timing information for the construction and evaluation stages.  Similar to automatic differentiation, the number of operations required to construct the AMI solution is nearly equivalent to the time required to evaluate the resulting solution a single time.   In Fig.~\ref{fig:hist}, we present the computational time in micro-seconds required, on average, to evaluate solutions to a single diagram of a given order.    To provide a fair representation we select specifically the diagrams involved in the self-energy expansion for a Hubbard interaction\cite{GIT,mcniven:2021}, and present the average time for evaluation per diagram across all diagrams at each order.  We see that even on a log scale there is a massive increase in computational expense with order which represents both a growth in the complexity of the integrand as well as the number of integration steps.  Again, the time required to construct the integrand has similar scaling, and is comparable to the timescale of a single evaluation of the internal variables.  Despite the large computational expense for higher order diagrams one should recognize that this is a small price to pay for the exact evaluation of a large fraction of the integration space (the temporal integrals) that for a problem in $d$ spatial dimensions represents a fraction $1/(d+1)$ of the integrals.  In addition, the computational expense of evaluation is not substantially larger than the evaluation of a hard-coded integrand, making the AMI construction extremely advantageous for high-order diagrams.

\section{License and citation policy}
The GitHub version of libami is licensed under the GNU General Public License
version 3 (GPL v. 3)\cite{gpl}. We kindly request that the present paper be cited, along with the original algorithmic paper \cite{AMI}, in any published work utilizing an application or code that uses this library.

\section{Summary}
We have presented a minimal framework for performing symbolic evaluation of Matsubara sums for arbitrary Feynman diagrams.  The core of \texttt{libami} is equivalent to a highly optimized symbolic math tool.  By avoiding use of generalized symbolic math packages, this AMI library can obtain solutions virtually instantaneously (on the scale of micro-seconds), as demonstrated in Section~\ref{sec:benchmarks}.  We have provided example content that should allow users to quickly integrate \texttt{libami} into existing Monte Carlo workflows.

\section{Acknowledgments}
We acknowledge funding from the Natural Sciences and 
Engineering Research Council of Canada grant RGPIN-2017-04253.

\appendix
\section{The SPR Evaluation}\label{app:ami}

To implement this procedure we define
the following objects:
\begin{itemize}
\item The arrays $R_p$ representing the configurations of Green's functions after the $p$th summation (described above).

\item The sets of poles $P_p$ for $\nu_p$ in the configuration of Green's functions represented by $R_{p-1}$.

\item The set of signs $S_p$ of the residues
for each pole. 
\end{itemize}
The array of poles corresponding to $\nu_p$ has entries
\begin{eqnarray}\label{E: poles_array}
P_p = [P_p^{(1)}, P_p^{(2)}, ..., P_p^{(r_{(p-1)})}],
\end{eqnarray}  
with each $P_p^{(\ell)}$ defined by a set (\texttt{std::vector}) of \texttt{pole\_struct}
\begin{eqnarray}\label{E: nu_p_poles_array}
P_p^{(\ell)} = [z_{p,\ell}^{(i_1)}, z_{p,\ell}^{(i_2)}, ..., z_{p,\ell}^{(i_{r_{\ell}} )}].
\end{eqnarray}  
We note that $P_p^{(\ell)}$ is the array of poles for $\nu_p$ in the residue of the $\ell$th pole for $\nu_{p-1}$ stored in the previous configuration of Green's functions, $R_{p-1}$. Similarly, we have an array of signs with the same dimensions as $P_p$,
\begin{eqnarray}\label{E: sign_array}
S_p = [S_p^{(1)}, S_p^{(2)}, ..., S_p^{(r_{(p-1)})}],
\end{eqnarray}  
with
\begin{eqnarray}\label{E: nu_s_poles_array}
S_p^{(\ell)} = [\alpha_{p,\ell}^{(i_1)}, \alpha_{p,\ell}^{(i_2)}, ..., \alpha_{p,\ell}^{(i_{r_{\ell}} )}],
\end{eqnarray}  
where $\alpha_{p,\ell}$ are the nonzero coefficients of $\nu_p$ from the previous configuration of the Green's functions, $R_{p-1}$.

Using these arrays, 
the full analytic result for Eq.~(\ref {eqn:goal}) is given by
\begin{eqnarray}\label{E: New_General_Sum_Again_Fianl_Again}
I^{(n)} = \frac{1}{\beta^n} \sum_{\{\nu_{n}\}} \prod_{j=1} ^ N G^j(\epsilon^j, X^j_m)  = K \cdot R_n,
\end{eqnarray}  
where
\begin{eqnarray}\label{E: Khorshid_Again}
K = (S_1 * f(P_1)) \times  ... \times (S_n * f(P_n)). \nonumber \\
\end{eqnarray}  
In this expression, $f(P_p)$ is the Fermi function of an array with elements
given by
\begin{eqnarray}\label{E: f_Op_new_compact}
[f(P_p)]_\ell^i = f(z_{p,\ell}^{(i)}),
\end{eqnarray}
and  the operations `$*$', `$\times$', and `$\cdot$' are defined by
\begin{eqnarray}\label{E: Dot_Gen_Compact}
(C * D)_i^j & = &  C_i^jD_i^j \equiv G_i^j, \nonumber \\ 
(G \times H)_i^j & =  & G_iH_i^j, \nonumber \\ 
H \cdot C  & = &  \sum_i H_i C_i. \nonumber 
\end{eqnarray}  
Equations (\ref{E: New_General_Sum_Again_Fianl_Again}) and (\ref{E: Khorshid_Again}) are primary objects of the evaluation stage of AMI. 
%

\section{Derivatives of Fermi Functions}\label{app:der}
We provide a single function \emph{fermi\_bose($M$,$\sigma$,$\beta$,$E$)} to generate the $M$'th derivative of Fermi or Bose functions ($\sigma=+1$ or $-1$, respectively) at inverse temperature $\beta$ and energy $E$. This is generated by a recursive procedure

\begin{eqnarray}
\frac{d^Mf(E)}{dE^M}=(-1)\beta^M\sum\limits_{k=0}^{M} F(M,k) \sigma^k(-1)^{k+1}\nonumber\\ \times \frac{1}{(\sigma \exp(\beta E)+1)}\frac{1}{(\sigma+\exp(-\beta E))^k }, \nonumber
\end{eqnarray}
where the $F$ function is a combinatorial prefactor.  We define the function
\begin{equation}
   F(r,k)=\sum\limits_{m=0}^{k}C(k,m) m^r(-1)^{k-m},
\end{equation}
where C(k,m) is the binomial coefficient.   

\bibliographystyle{elsarticle-num}
\bibliography{refs.bib}

\end{document}